\documentclass[aps, pre, superscriptaddress, twocolumn, showpacs, amsmath, longbibliography, floatfix]{revtex4-1}

\usepackage{dcolumn}
\usepackage{bm} 
\usepackage{graphicx} 
\usepackage{color}
\usepackage{subfigure}
\usepackage{hyperref}
\usepackage{latexsym}
\usepackage{amsthm}
\usepackage{amssymb}
\usepackage{verbatim}
\usepackage{wasysym}

\DeclareGraphicsExtensions{.jpg, .pdf, .mps, .png, .eps, .ps, .EPS, .gif}
\begin{document}
\def\be{\begin{equation}}
\def\ee{\end{equation}}

\def\bc{\begin{center}}
\def\ec{\end{center}}
\def\bea{\begin{eqnarray}}
\def\eea{\end{eqnarray}}

\newcommand{\avg}[1]{\langle{#1}\rangle}
\newcommand{\Avg}[1]{\left\langle{#1}\right\rangle}

\newcommand{\cch}[1]{\left[#1\right]}
\newcommand{\chv}[1]{\left \{ #1\right \} }
\newcommand{\prt}[1]{\left(#1\right)}
\newcommand{\aver}[1]{\left\langle #1 \right\rangle}
\newcommand{\abs}[1]{\left| #1 \right|}

\def\ie{\textit{i.e.}}
\def\etal{\textit{et al.}}
\def\m{\vec{m}}
\def\G{\mathcal{G}}
\def\fig{Fig.}
\def\tab{Table }

\newcommand{\red}[1]{{#1}}
\newcommand{\redChangesResub}[1]{{\color{red}#1}}

\title{Precise bond percolation thresholds on several four-dimensional lattices}

\author{Zhipeng Xun}
\email{zpxun@cumt.edu.cn}
\affiliation{School of Physical Science and Technology, China University of Mining and Technology, Xuzhou 221116, China}
\author{Robert M. Ziff}
\email{rziff@umich.edu}
\affiliation{Center for the Study of Complex Systems and 
Department of Chemical Engineering, University of Michigan, 
Ann Arbor, Michigan 48109-2136, USA}

\date{\today}

\begin{abstract} 
We study bond percolation on several four-dimensional (4D) lattices, including the simple (hyper) cubic (SC), the SC with combinations of nearest neighbors and second nearest neighbors (SC-NN+2NN), the body-centered cubic (BCC), and the face-centered cubic (FCC) lattices, using an efficient single-cluster growth algorithm. For the SC lattice, we find $p_c = 0.1601312(2)$, which confirms previous results (based on other methods), and find a new value $p_c=0.035827(1)$ for the SC-NN+2NN lattice, which was not studied previously for bond percolation. For the 4D BCC and FCC lattices, we obtain $p_c=0.074212(1)$ and 0.049517(1), which are substantially more precise than previous values.  We also find critical exponents $\tau = 2.3135(5)$ and $\Omega = 0.40(3)$, consistent with previous numerical results and the recent four-loop series result of Gracey [Phys.\ Rev.\ D {\bf 92}, 025012, (2015)].
\end{abstract}

\pacs{64.60.ah, 89.75.Fb, 05.70.Fh}

\maketitle
\section{Introduction}
Percolation, which was introduced by Broadbent and Hammersley  \cite{BroadbentHammersley1957} in 1957,
is one of the fundamental models in statistical physics \cite{StaufferAharony1994,Grimmett1999}. In percolation systems, sites or bonds on a lattice are either occupied with probability $p$, or not with probability $1-p$. When increasing $p$ from below, a cluster large enough to span the entire system from one side to the other will first appear at a value $p_{c}$. This point is called the percolation threshold.

The percolation threshold is an important physical quantity, because many interesting phenomena, such as phase transitions, occur at that point. Consequently, finding percolation thresholds for a variety of lattices has been a long-standing subject of research in this field. In two dimensions, percolation thresholds of many lattices can be found analytically  \cite{SykesEssam1964,Scullard2006,Ziff2006,ZiffScullard2006}, while others must be found numerically. In three and higher dimensions, there are no exact results, and all thresholds must be determined by approximation schemes or numerical methods. Many effective numerical simulation algorithms \cite{HoshenKopelman1976,Leath1976,NewmanZiff2000,NewmanZiff2001} have been developed. For example, the ``cluster multiple labeling technique" was proposed by Hoshen and Kopelman \cite{HoshenKopelman1976} to determine the critical percolation concentration, percolation probabilities, and cluster-size distributions for percolation problems. Newman and Ziff  \cite{NewmanZiff2000,NewmanZiff2001} developed a Monte Carlo algorithm which allows one to calculate quantities such as the cluster-size distribution or spanning probability over the entire range of site or bond occupation probabilities from zero to one in a single run, and takes an amount of time that scales roughly linearly with the number of sites on the lattice.  

Much work in finding thresholds has been done with these and other techniques. Series estimates of the critical percolation probabilities for the bond problem and the site problem were presented by Sykes and Essam \cite{SykesEssam1964-2}, which can be traced back to 1960s. Lorenz and Ziff \cite{LorenzZiff1998} performed extensive Monte Carlo simulations to study bond percolation on three-dimensional lattices using an epidemic cluster-growth approach. Determining the crossing probability  \cite{ReynoldsStanleyKlein80,StaufferAharony1994,FumikoShoichiMotoo1989} $R(p)$ as a function of $p$ for different size systems, and using scaling to analyze the results is also a common way to find $p_{c}$. \red{Binder ratios have also been used to determine the threshold \cite{WangZhouZhangGaroniDeng2013,Norrenbrock16,SampaioFilhoCesarAndradeHerrmannMoreira18}.} By examining wrapping probabilities, Wang et al.\ \cite{WangZhouZhangGaroniDeng2013} and Xu et al.\ \cite{XuWangLvDeng2014} simulated the bond and site percolation models on several three-dimensional lattices, including simple cubic (SC), the diamond, body-centered cubic (BCC), and face-centered cubic (FCC) lattices.  Other recent work on percolation includes Refs.\ \cite{MitraSahaSensharma19,KryvenZiffBianconi19,RamirezCentresRamirezPastor19,GschwendHerrmann19,Koza19,MertensMoore2018,MertensMoore18s,HassanAlamJituRahman17,KennaBerche17,MertensJensenZiff17}.

Percolation has been investigated on many kinds of lattices. In three and higher dimensions, the most common of these lattices are the SC, the BCC, and the FCC lattices. Thanks to the techniques mentioned above, precise estimates are known for the critical thresholds for site and bond percolation and related exponents in three dimensions. However, in four dimensions (4D), the estimates of bond percolation thresholds that have been determined for the BCC and FCC lattices are much less precise \cite{vanderMarck98} than the values that have been found for some other lattices \red{(that is, two vs.\ five or six significant digits).} In addition, to the best of our knowledge, the bond percolation threshold on SC lattice with the combinations of nearest neighbors (NN) and second nearest neighbors (2NN), namely (SC-NN+2NN), has not been reported so far.  We note that the notation 2N+3N is also used for NN and 2NN \cite{MalarzGalam05}.

In this paper, we employ the single-cluster growth method \cite{LorenzZiff1998} to study bond percolation on several lattices in 4D. While confirming previous results of SC lattice, we obtain more precise estimates of percolation thresholds for BCC and FCC lattices. We also find a new value for bond threshold of the complex-neighborhood lattice, SC-NN+2NN.  \red{Note that percolation on lattices with complex neighborhoods can also be interpreted as the percolation of extended objects on a lattice that touch or overlap each other \cite{KozaKondratSuszczynski14}.} 

With regards to the latter system, Malarz and co-workers \cite{MalarzGalam05,MajewskiMalarz2007,KurzawskiMalarz2012,Malarz2015,KotwicaGronekMalarz19} have carried out several studies on lattices with various complex neighborhoods, that is, lattices with combinations of two or more types of neighbor connections, in two, three and four dimensions. Their results have all concerned site percolation, and are generally given to only three significant digits.  Here we show that the single-cluster growth method can be efficiently applied to one of these lattices also.  Our goal was  to find results to at least five significant digits, which was not difficult to achieve using the methods given here.  \red{Note that in general, for Monte Carlo work, increasing the precision by one digit requires at least 100 times more work in terms of the number of simulations, not to mention the additional work studying corrections to scaling and other necessary quantities.}

Precise percolation thresholds are needed in order to study the critical behavior, including critical exponents, critical crossing probabilities, critical and excess cluster numbers, etc.  Four dimensions is interesting because it is close enough to six dimensions for $6-\epsilon$ series analysis to have a hope of yielding good results \cite{Gracey2015}, and in general there is interest on how thresholds depend upon dimensionality \cite{GauntRuskin78,vanderMarck98,vanderMarck98k,Grassberger03,TorquatoJiao13,MertensMoore2018}.  The study of how thresholds depend upon lattice structure, especially the coordination number $z$, has also had a long history \cite{ScherZallen70,GalamMauger96,vanderMarck97,Wierman02,WiermanNaor05}.  Having thresholds of more lattices is useful for extending those correlations.

In the following sections, we present the underlying theory, and discuss the simulation process. Then we present and briefly discuss the results that we obtained from our simulations.

\section{Theory}\label{sec:model}
The central property describing the cluster statistics in percolation is $n_{s}$, defined as the number of clusters (per site) containing $s$ occupied sites or bonds, as a function of the occupation probability $p$. At the percolation threshold $p_{c}$, $n_{s}$ is expected to behave as

\begin{equation}
n_{s} \sim A_0 s^{-\tau} (1+B_0 s^{-\Omega}+\dots),
\label{eq:ns}
\end{equation}
where $\tau$ is the Fisher exponent, and $\Omega$ is the leading correction-to-scaling exponent.  Both $\tau$ and $\Omega$ are expected to be universal, namely the same for all lattices of a given dimensionality. The $A_0$ and $B_0$ are constants that depend upon the system (are non-universal). The probability a vertex belongs to a cluster with size greater than or equal to $s$ will then be
\begin{equation}
P_{\geq s} = \sum_{s'=s}^\infty s' n_{s'} \sim A_1s^{2-\tau} (1+B_1s^{-\Omega}+\dots),
\label{ps}
\end{equation}
where $A_1 = A_0/(\tau-2)$ and $B_1 = (\tau-2)B_0/(\tau+\Omega-2)$.  Multiplying both sides of Eq.\  (\ref{ps}) by $s^{\tau-2}$, we have
\begin{equation}
s^{\tau-2}P_{\geq s} \sim A_1 (1+B_1 s^{-\Omega}+\dots).
\label{staup}
\end{equation}
It can be seen that there will be a linear relationship between $s^{\tau-2}P_{\geq s}$ and $s^{-\Omega}$ for large $s$, if we choose the correct value of $\Omega$. This linear relationship can be used to determine the value of percolation threshold, because for $p \ne p_c$ the behavior  will be nonlinear.

Taking the logarithm of Eq.\  (\ref{ps}), we find
\begin{equation}
\begin{aligned}
\ln P_{\geq s} & \sim \ln A_1 + (2-\tau)\ln s + \ln(1+B_1 s^{-\Omega}) \\
& \sim \ln A_1 + (2-\tau)\ln s + B_1 s^{-\Omega},
\end{aligned}
\end{equation}
for large $s$. Similarly,
\begin{equation}
\ln  P_{\geq 2s} \sim \ln A_1 + (2-\tau)\ln 2 s + B_1(2s)^{-\Omega}.
\end{equation}
Then it follows that
\begin{equation}
\begin{aligned}
\frac{\ln P_{\geq 2s} - \ln P_{\geq s}}{\ln 2} &\sim \frac{(2 - \tau)(\ln 2s - \ln s)}{\ln 2} - \frac{B_1 s^{-\Omega}(2^{-\Omega}-1)}{\ln 2} \\
&\sim (2 - \tau) + B_2 s^{-\Omega} ,
\end{aligned}
\label{localslope}
\end{equation}
where $(\ln P_{\geq 2s} - \ln P_{\geq s})/\ln 2$ is  the local slope of a plot of $\ln P_{\geq 2s}$ vs.\ $\ln s$, and $B_2 = B_1(2^{-\Omega}-1)/\ln 2$.
Eq.\ (\ref{localslope}) implies that if we make of plot of the local slope vs.\  $s^{-\Omega}$ at $p_c$, linear behavior will be found for large $s$, and the intercept of the straight line will give the value of $(2-\tau)$.  \red{Of course, there will be higher-order corrections to Eqs.\ (\ref{eq:ns}) and (\ref{localslope}) related
to an (unknown) exponent $\Omega_1$, but for large $s$ linear behavior in this plot should be found.  We did not attempt to characterize the higher-order corrections to scaling.}

\section{Simulation results and discussions}
The basic algorithm of single-cluster growth method is as follows. An individual cluster starts to grow at the seeded site that is located on the lattice. We choose the origin of coordinates for the seeded site,  though any site on the lattice can be chosen under periodic \red{or helical} boundary conditions. From this site, a cluster is grown to neighboring sites by occupying the connecting bonds with a certain probability $p$ or leaving them unoccupied with probability $1-p$.  All of these clusters are allowed to grow until they terminate in a complete cluster, or when they reach an upper size cutoff, their growing is halted.

To grow the clusters, we check all neighbors of a growth site for unvisited sites, which we occupy with probability $p$, and put the newly occupied growth site on a first-in, first-out queue. \red{Growth sites are those occupied whose neighbors have yet to be checked for further growth, and unvisited sites are those site whose occupation has not yet been determined, for one particular run.}  To simulate bond percolation, we simply leave the sites in the unvisited state when we do not occupy them through an occupied bond.  (For site percolation, unoccupied visited sites are blocked from ever being occupied in the future.)  The single-cluster growth method is similar to the Leath algorithm \cite{Leath1976}. 

We utilize a simple programming procedure to avoid clearing out the lattice after each cluster is formed: the lattice values are started out at $0$, and for cluster (run) $n$, any site whose value is less than $n$ is considered unoccupied.  When a site is occupied in the growth of a new cluster, it is assigned the value $n$.  This procedure saves a great deal of time because we can use a very large lattice, and do not have to clear out the whole lattice after every cluster, many of which are small.

\red{Following is a pseudo-code of this basic algorithm:}

\medskip

\red{\tt Set all lat[x] = 0

for runs = 1 to runsmax

\quad Put origin on queue

\quad set lat[0]=runs

\quad do  

\quad \quad get x = oldest member of queue

\quad \quad for dir = 0 to directionmax-1

\quad \quad  \quad  set xp = x + deltax[dir]
   
\quad \quad \quad  if (lat[xp \& W] < runs)
   
\quad \quad  \quad \quad if (rnd < prob)

\quad \quad \quad \quad  \quad set lat[xp \& W] = runs

\quad \quad \quad \quad  \quad put xp on queue

\quad while ((queue != empty) \&\& (size < max))
   
}
 
\medskip      
   
\red{The actual code is not too many lines longer than this.  {\tt rnd} is a random in the range $(0,1)$.  {\tt max} is the maximum cluster size, $2^{15}$ to $2^{17}$ here.  We use a one-dimensional array {\tt lat} of length $L^4 = 2^{28} = 268435456$, and use the 
bit-wise ``and" function {\tt \&} to carry out the helical wraparound by writing {\tt lat[xp \& W]} with $W = L^4 - 1$.  (This works only for $L$ that are powers of two.)  The {\tt deltax} array are the eight values $1, -1, L, -L, L^2, - L^2, L^3, -L^3$ for the SC lattice, and generalized accordingly for the other lattices.  The size of the cluster is just the value of the queue insert pointer.}

\red{For site percolation, one simply replaces the last five lines by 

{\tt \quad \quad \quad  if (lat[xp \& W] < runs)
   
\quad \quad \quad \quad set lat[xp \& W] = runs

\quad \quad  \quad  \quad  if (rnd < prob)

\quad \quad \quad \quad \quad put xp on queue

\quad while ((queue != empty) \&\& (size < max))}

}

The size of the cluster is identified by the number of occupied sites it contains. Then the number of clusters whose size (number of sites) fall in a range of $(2^n, 2^{n+1}-1)$ for $n=0,1,2,\cdot\cdot\cdot$ is recorded in the $n$th bin. If a cluster is still growing when it reaches the upper cutoff, it is counted in the last bin. The cutoff was $2^{17}$ occupied sites for the SC lattice, $2^{16}$ for FCC and SC-NN+2NN, and $2^{15}$ for the BCC lattice.  The cutoff had to be lower in the latter case because of the expanded nature of the BCC lattice represented on the SC lattice.

While the single-cluster growth method requires separate runs to be made for different values of $p$, it is not difficult to quickly zero in on the threshold to four or five digits, and then reserve the longer runs for finding the sixth digit.  It is also simple to analyze the results as shown here --- one does not need to study things like the intersections of crossing probabilities for different size systems or create large output files of intermediate microcanonical results to find estimates of the threshold.  The output files here are simply the 15 to 17 values of the bins for each value of $p$ described above.

The simulations on the SC lattice, SC-NN+2NN lattice, BCC lattice, and FCC lattice were carried out for system size $L\times L \times L \times L$ with $L=128$, and with periodic boundary conditions. For each lattice, we produced $10^9$ independent samples. Then the number of clusters greater than or equal to size $s$ could be found based on the data from our simulation, and the corresponding quantities, such as the local slope $((\ln P_{\geq 2s} - \ln P_{\geq s})/\ln 2)$, and $s^{\tau-2}P_{\geq s}$, could be easily calculated.

\begin{figure}[htbp] 
\centering
\includegraphics[width=3.8in]{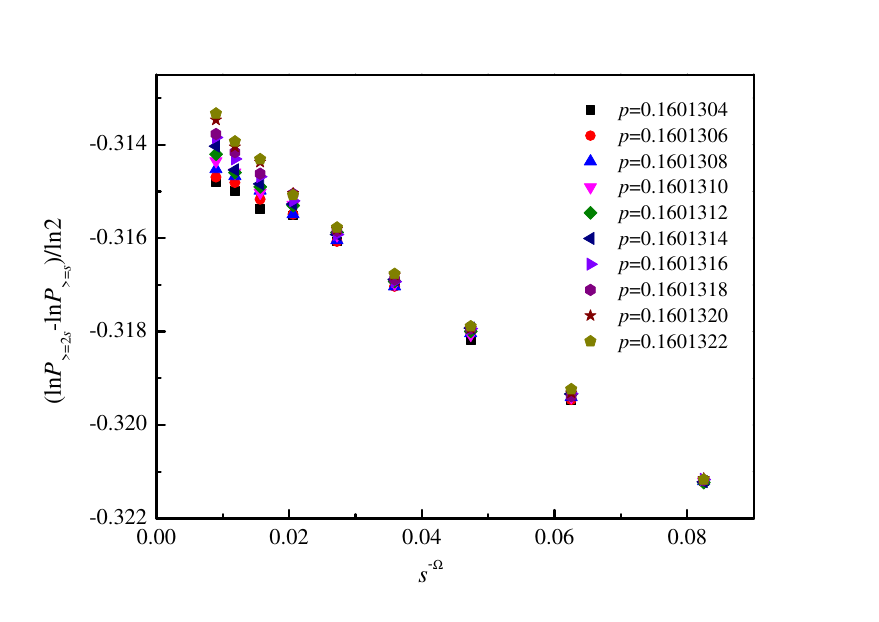} 
\caption{Plot of the local slope $(\ln P_{\geq 2s} - \ln P_{\geq s})/\ln 2$ vs.\ $s^{-\Omega}$ \red{with $\Omega = 0.40$} for the SC lattice under different values of $p$. \red{The solid line in the figure is a guideline through the data points for $p = 0.1601312 \approx p_c$. The intercept -0.3135 is an estimate for $2 - \tau$ by Eq.\ (\ref{localslope}).}}
\label{fig:sc-localslope-vs-s-omega}
\end{figure}

\begin{figure}[htbp] 
\centering
\includegraphics[width=3.8in]{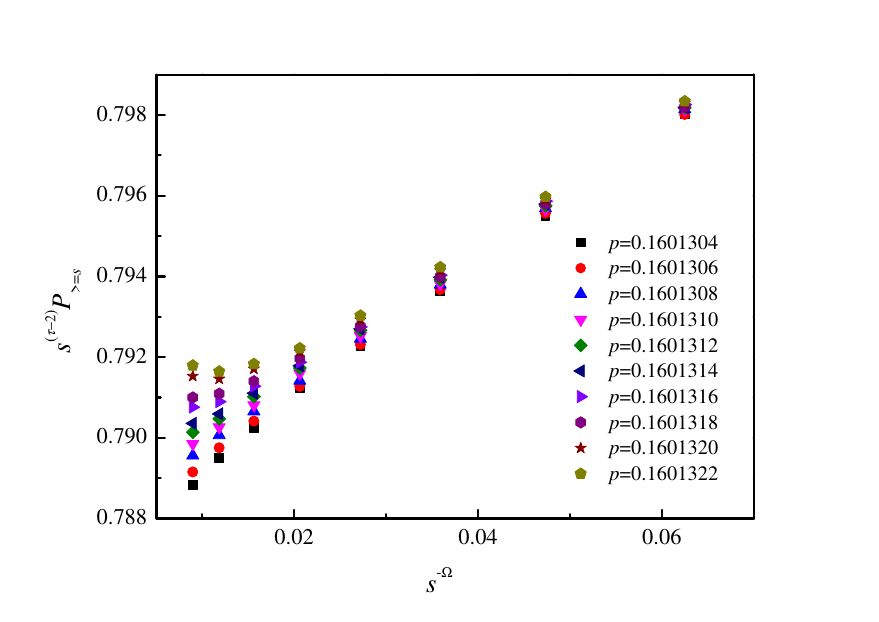} 
\caption{Plot of $s^{\tau-2}P_{\geq s}$ vs.\ $s^{-\Omega}$ \red{with $\Omega = 0.40$} for the SC lattice under different values of $p$. \red{The solid line in the figure is a guideline following the points for $p = 0.1601312 \approx p_c$.}}
\label{fig:sc-s-tau-2-Ps-vs-s-omega}
\end{figure}

Figs.\ \ref{fig:sc-localslope-vs-s-omega} and \ref{fig:sc-s-tau-2-Ps-vs-s-omega}, respectively, show the plots of the local slope and $s^{\tau-2}P_{\geq s}$ vs.\ $s^{-\Omega}$ for the SC lattice under different values of $p$. When $p$ is away from $p_{c}$, no matter if it is larger or smaller than $p_{c}$, the curves show a deviation from linearity. When $p$ is very near to $p_{c}$, we can see better linear behavior for large $s$. The linear behavior here is in good agreement with the theoretical predictions of Eqs.\ (\ref{staup}) and (\ref{localslope}).

Based on these simulation results, for bond percolation on the SC lattice in 4D, we conclude

~\\
SC:

$p_c = 0.1601312(2)$,   $\tau = 2.3135(7)$, and $\Omega = 0.40(3)$.

~\\
Here numbers in parentheses represent errors in the last digit(s), determined from the observed statistical errors.

\begin{figure}[htbp] 
\centering
\includegraphics[width=3.8in]{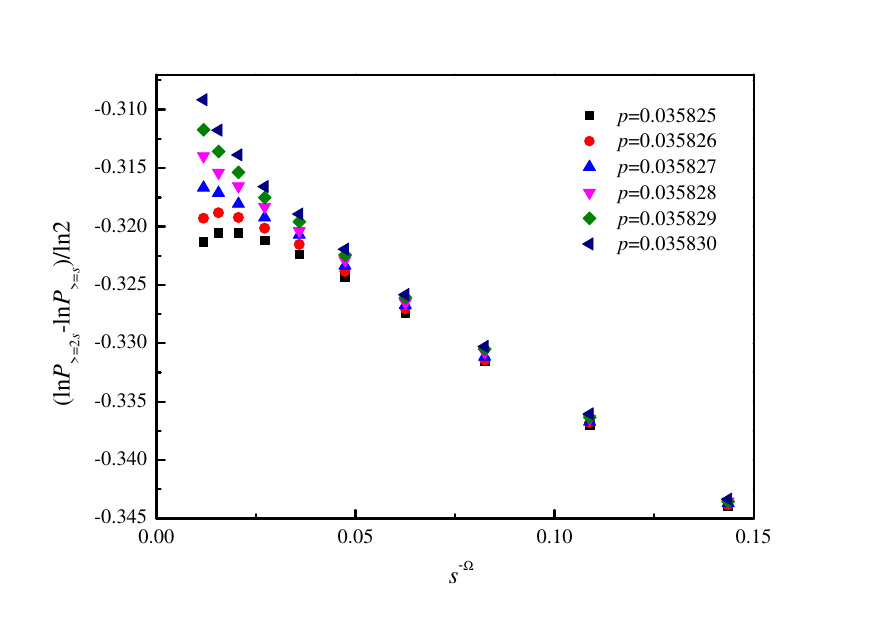} 
\caption{Plot of the local slope $((\ln P_{\geq 2s} - \ln P_{\geq s})/\ln 2)$ vs.\ $s^{-\Omega}$ \red{with $\Omega = 0.40$} for the SC-NN+2NN lattice under different values of $p$. \red{The solid line in the figure is a guideline through the data points for $p = 0.035827 \approx p_c$. The intercept -0.3137 is an estimate for $2 - \tau$ by Eq.\ (\ref{localslope}).}}
\label{fig:sc-NN2NN-localslope-vs-s-omega}
\end{figure}

\begin{figure}[htbp] 
\centering
\includegraphics[width=3.8in]{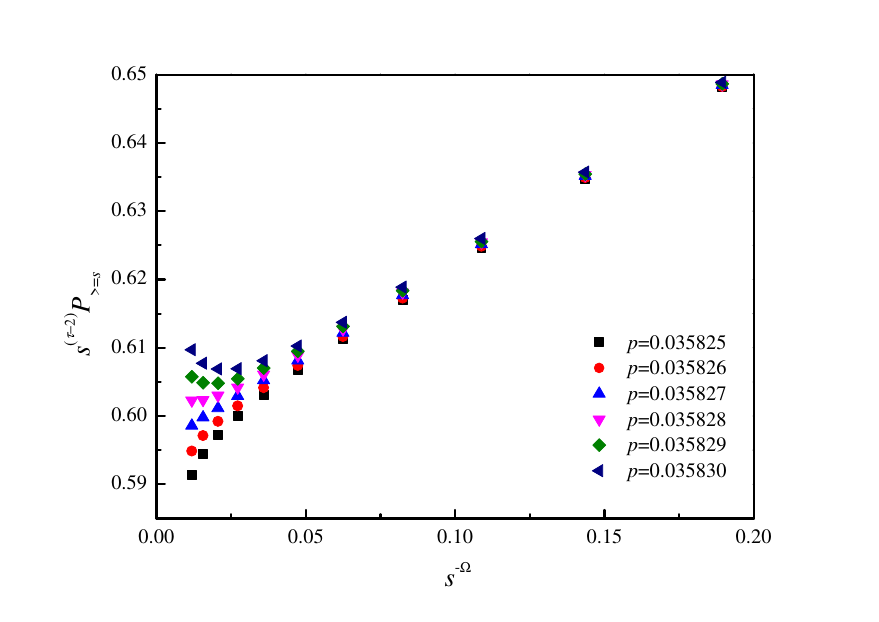} 
\caption{Plot of $s^{\tau-2}P_{\geq s}$ vs.\ $s^{-\Omega}$ \red{with $\Omega = 0.40$} for the SC-NN+2NN lattice under different values of $p$. \red{The solid line in the figure is a guideline following the points for $p = 0.035827 \approx p_c$.}}
\label{fig:sc-NN2NN-s-tau-2-Ps-vs-s-omega}
\end{figure}

\begin{figure}[htbp] 
\centering
\includegraphics[width=3.8in]{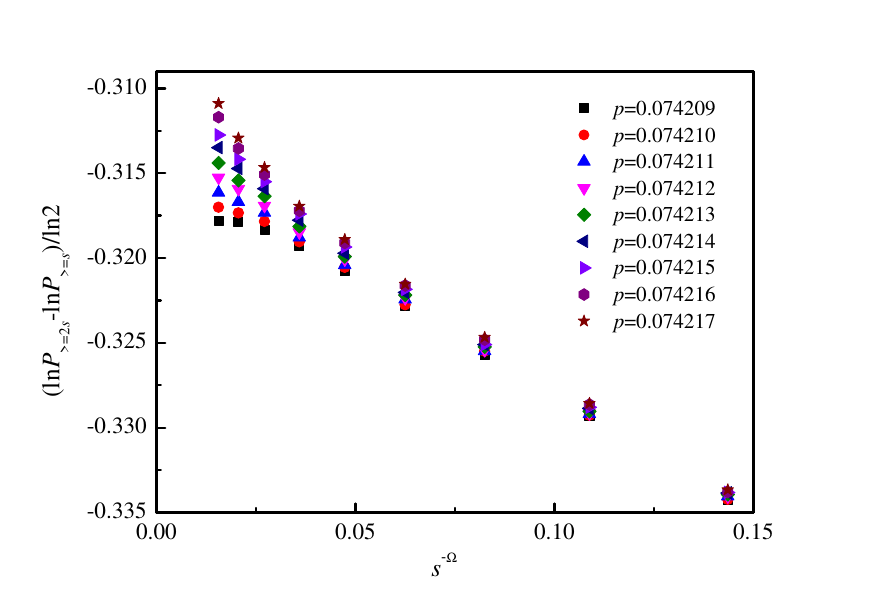} 
\caption{Plot of the local slope $((\ln P_{\geq 2s} - \ln P_{\geq s})/\ln 2)$ vs.\ $s^{-\Omega}$ \red{with $\Omega = 0.41$} for the BCC lattice under different values of $p$. \red{The solid line in the figure is a guideline through the data points for $p = 0.074212 \approx p_c$. The intercept -0.3134 is an estimate for $2 - \tau$ by Eq.\ (\ref{localslope}).} }
\label{fig:bcc-localslope-vs-s-omega}
\end{figure}

\begin{figure}[htbp] 
\centering
\includegraphics[width=3.8in]{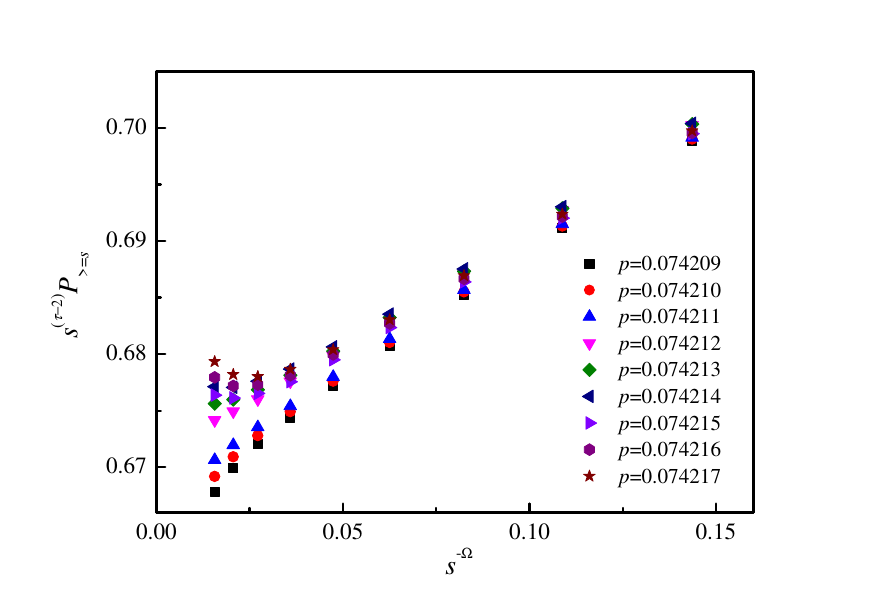} 
\caption{Plot of $s^{\tau-2}P_{\geq s}$ vs.\ $s^{-\Omega}$ \red{with $\Omega = 0.41$} for the BCC lattice under different values of $p$. \red{The solid line in the figure is a guideline following the points for $p = 0.074212 \approx p_c$.}}
\label{fig:bcc-s-tau-2-Ps-vs-s-omega}
\end{figure}

\begin{figure}[htbp] 
\centering
\includegraphics[width=3.8in]{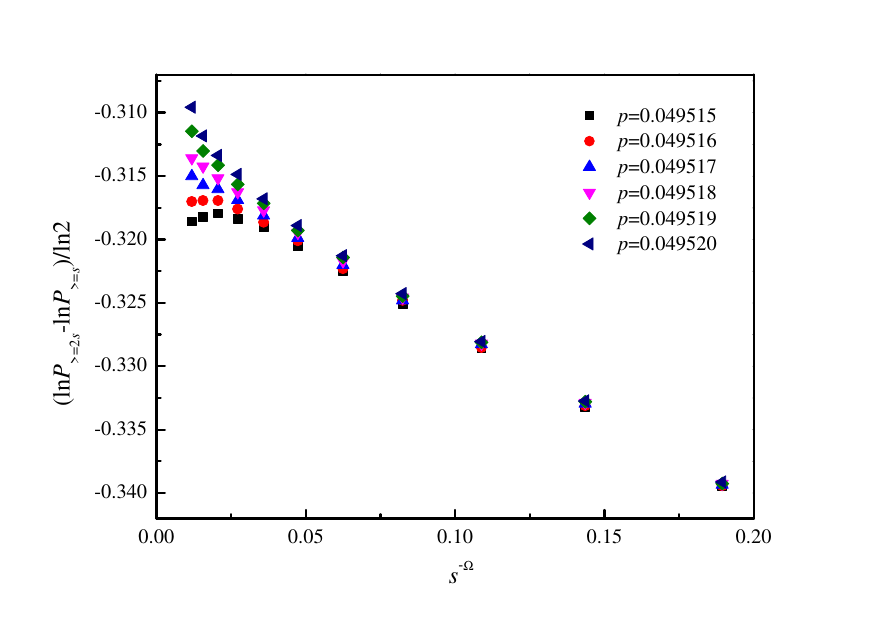} 
\caption{Plot of the local slope $((\ln P_{\geq 2s} - \ln P_{\geq s})/\ln 2)$ vs.\ $s^{-\Omega}$ \red{with $\Omega = 0.41$} for the FCC lattice under different values of $p$. \red{The solid line in the figure is a guideline through the data points for $p = 0.049517 \approx p_c$. The intercept -0.3135 is an estimate for $2 - \tau$ by Eq.\ (\ref{localslope}).}}
\label{fig:fcc-localslope-vs-s-omega}
\end{figure}

\begin{figure}[htbp] 
\centering
\includegraphics[width=3.8in]{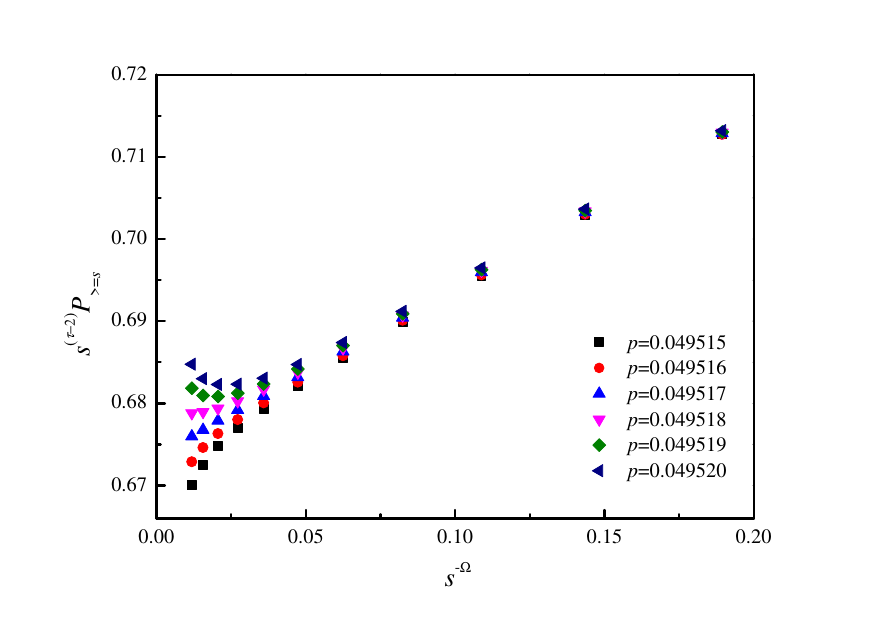} 
\caption{Plot of $s^{\tau-2}P_{\geq s}$ vs.\ $s^{-\Omega}$ \red{with $\Omega = 0.41$} for the FCC lattice under different values of $p$. \red{The solid line in the figure is a guideline following the points for $p = 0.049517 \approx p_c$.}}
\label{fig:fcc-s-tau-2-Ps-vs-s-omega}
\end{figure}

The simulation results for other three lattices, i.e., the plots of the local slope and $s^{\tau-2}P_{\geq s}$ vs.\ $s^{-\Omega}$ for the SC-NN+2NN, BCC, and FCC lattices under different values of $p$ are shown in Figs.\ \ref{fig:sc-NN2NN-localslope-vs-s-omega},\ \ref{fig:sc-NN2NN-s-tau-2-Ps-vs-s-omega}, \ref{fig:bcc-localslope-vs-s-omega}, \ref{fig:bcc-s-tau-2-Ps-vs-s-omega}, \ref{fig:fcc-localslope-vs-s-omega}, and \ref{fig:fcc-s-tau-2-Ps-vs-s-omega}. From these figures, we can see similar behavior as the SC lattice. In order to avoid unnecessary repetition, we do not  discuss the data one by one, and directly show the deduced values of $p_{c}$ and the two exponents below.

~\\
SC-NN+2NN:

$p_{c} = 0.035827(1)$, $\tau = 2.3138(12)$, and $\Omega = 0.40(3)$.

~\\
BCC:

$p_{c} = 0.074212(1)$, $\tau = 2.3133(9)$, and $\Omega = 0.41(3)$.

~\\
FCC:

$p_{c} = 0.049517(1)$, $\tau = 2.3135(9)$, and $\Omega = 0.41(3)$.
~\\

From these values, we have obtained precise estimates of the percolation threshold, and also confirmed the universality of the Fisher exponent $\tau$. 

When the probability $p$ is away from $p_{c}$, a scaling function needs to be included. Then the behavior can be represented as 
\begin{equation}
P_{\geq s} \sim A_2 s^{2-\tau} f(B_2(p-p_{c})s^{\sigma}),
\label{ps2}
\end{equation}
in the scaling limit of $s \rightarrow \infty$ and $p \rightarrow p_{c}$. The scaling function $f(x)$ can be expanded as a Taylor series,

\begin{equation}
f(B_2(p-p_{c})s^{\sigma}) \sim 1+C_2(p-p_{c})s^{\sigma}+ \cdot\cdot\cdot.
\label{scaling}
\end{equation}
where $C_2 = B_2 f'(0)$.  We assume $f(0)=1$, so that $A_2$ = $A_1$.

\begin{figure}[htbp] 
\centering
\includegraphics[width=3.8in]{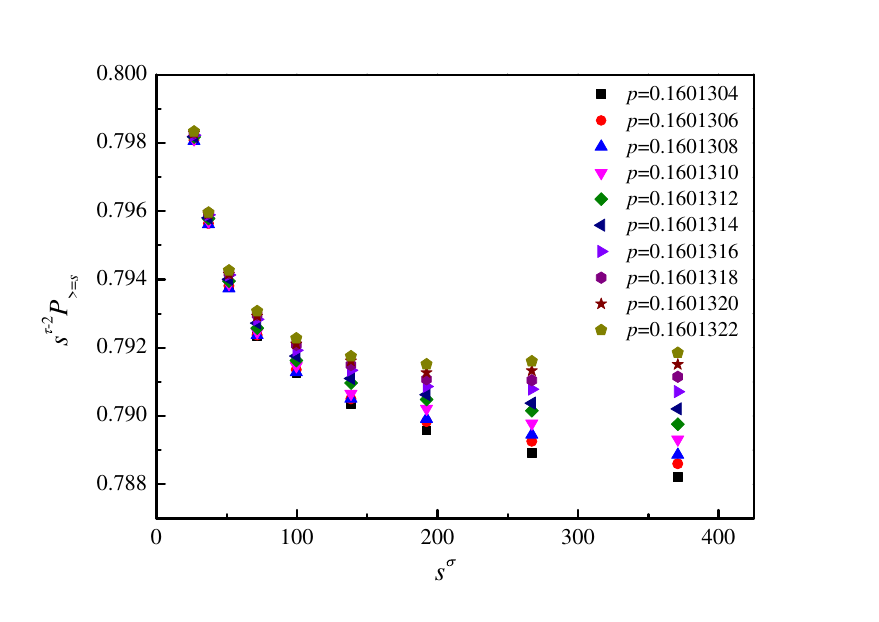} 
\caption{Plot of $s^{\tau-2}P_{\geq s}$ vs.\ $s^{\sigma}$ \red{with $\sigma=0.4742$ and $\tau = 2.3135$} for the SC lattice under different values of $p$. \red{The dashed line in the figure is a guideline through the points for $p = 0.1601314 \approx p_c$}.}
\label{fig:sc-s-tau-2-Ps-vs-s-sigma}
\end{figure}

\begin{figure}[htbp] 
\centering
\includegraphics[width=3.8in]{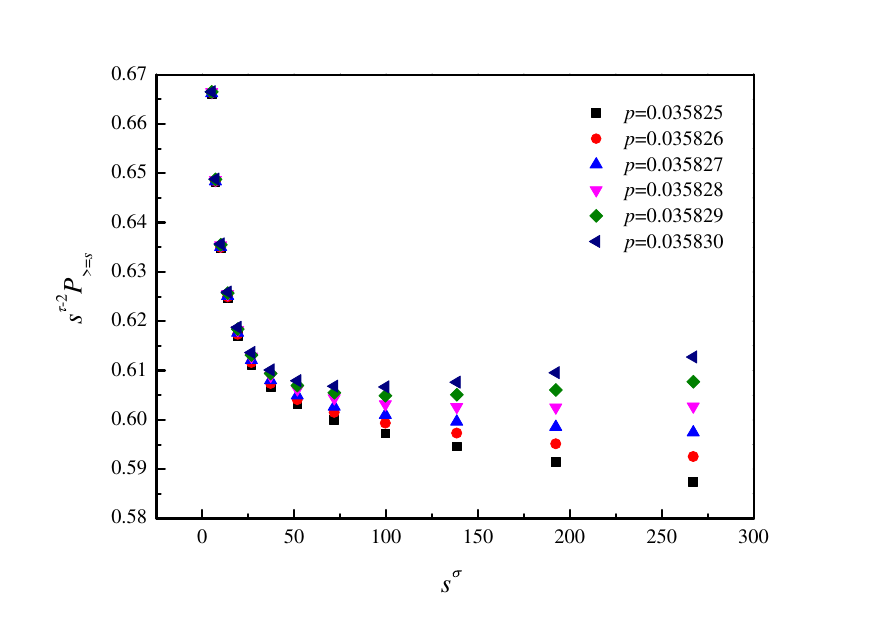} 
\caption{Plot of $s^{\tau-2}P_{\geq s}$ vs.\ $s^{\sigma}$ \red{with $\sigma=0.4742$ and $\tau = 2.3137$} for the SC-NN+2NN lattice under different values of $p$. \red{ The dashed line in the figure is a guideline through the points for $p = 0.035827 \approx p_c$}.}
\label{fig:sc-NN2NN-s-tau-2-Ps-vs-s-sigma}
\end{figure}

\begin{figure}[htbp] 
\centering
\includegraphics[width=3.8in]{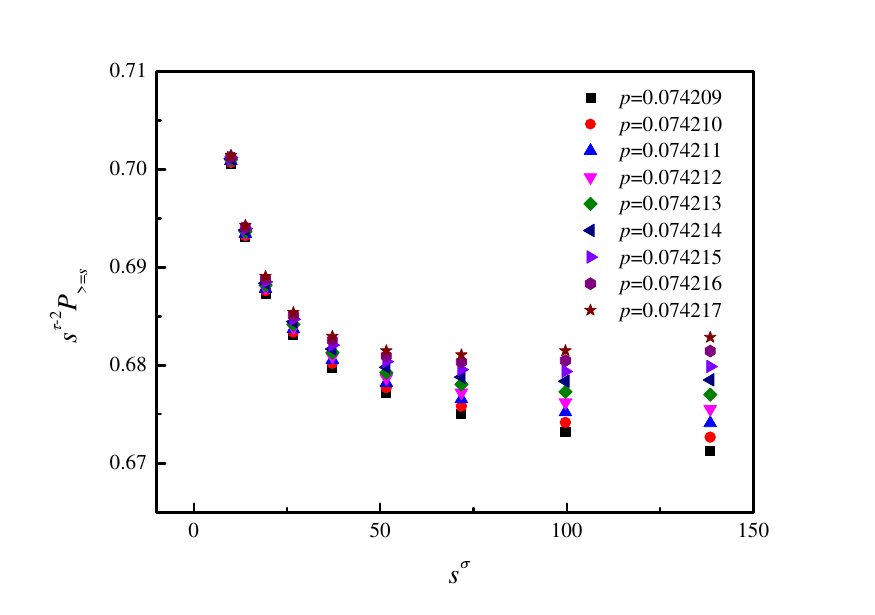} 
\caption{Plot of $s^{\tau-2}P_{\geq s}$ vs.\ $s^{\sigma}$ \red{with $\sigma=0.4742$ and $\tau = 2.3134$} for the BCC lattice under different values of $p$. \red{The dashed line in the figure is a guideline through the points for $p = 0.074212 \approx p_c$}.}
\label{fig:bcc-s-tau-2-Ps-vs-s-sigma}
\end{figure}

\begin{figure}[htbp] 
\centering
\includegraphics[width=3.8in]{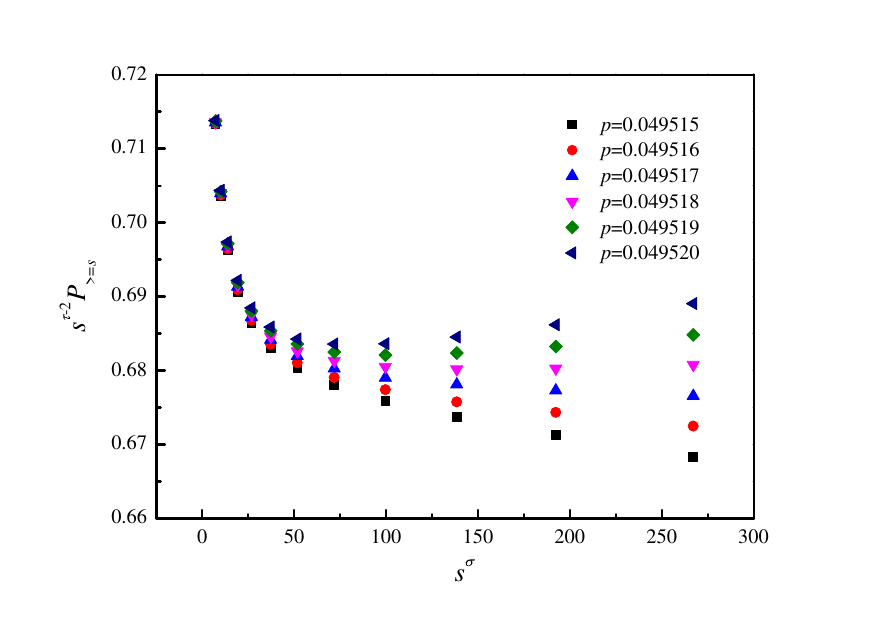} 
\caption{Plot of $s^{\tau-2}P_{\geq s}$ vs.\ $s^{\sigma}$ \red{with $\sigma=0.4742$ and $\tau = 2.3135$} for the FCC lattice under different values of $p$. \red{The dashed line in the figure is a guideline through the points for $p = 0.049517 \approx p_c$.}}
\label{fig:fcc-s-tau-2-Ps-vs-s-sigma}
\end{figure}

Combining Eqs.\ (\ref{ps2}) and (\ref{scaling}) leads to
\begin{equation}
s^{\tau-2}P_{\geq s} \sim A_2+D_2(p-p_{c})s^{\sigma}.
\label{vssigma}
\end{equation}
where $D_2=A_2 C_2$. 
Eq.\ (\ref{vssigma}) predicts that $s^{\tau-2}P_{\geq s}$ will convergence to a constant value at $p_{c}$ for large $s$, while it deviates from a constant value when $p$ is away from $p_{c}$. This provides another way to determine the percolation threshold. Figs.\ \ref{fig:sc-s-tau-2-Ps-vs-s-sigma}, \ref{fig:sc-NN2NN-s-tau-2-Ps-vs-s-sigma}, \ref{fig:bcc-s-tau-2-Ps-vs-s-sigma}, \ref{fig:fcc-s-tau-2-Ps-vs-s-sigma} show the plots of $s^{\tau-2}P_{\geq s}$ versus $s^{\sigma}$ for the SC, SC-NN+2NN, BCC and FCC lattices, respectively. For these plots, we use the value of $\sigma=0.4742$, which is provided in Ref.\  \cite{Gracey2015}. The estimations of percolation thresholds are shown below, and they are consistent with the values obtained above.
~\\

SC: $p_{c} = 0.1601314(2)$.
~\\

SC-NN+2NN: $p_{c} = 0.035827(1)$.
~\\

BCC: $p_{c} = 0.074212(1)$.
~\\

FCC: $p_{c} = 0.049517(1)$.
~\\

Our final estimates of percolation thresholds for all the lattices calculated in this paper are summarized in Table \ref{tab:ept}, where we also make a comparison with those of previous studies where available. It can be seen that for the SC lattice, our result is completely consistent with the existing ones within the error range, including the recent more precise result of Mertens and Moore \cite{MertensMoore2018}.

\red{To find our results for $p_c$, $\tau$ and $\Omega$, we basically adjusted these values to get the best linear behavior on the two plots
of $(\ln P_{\geq 2s} - \ln P_{\geq s})/\ln 2$ vs.\ $s^{-\Omega}$ and $s^{\tau-2}P_{\geq s}$ vs.\ $s^{-\Omega}$,  and horizontal asymptotic behavior on the plot of $s^{\tau-2}P_{\geq s}$ vs.\ $s^{\sigma}$, for each of the four lattices.  With the incorrect value of $\Omega$, for example, we would not get linear behavior over several orders of magnitude of $s$ for any value of $p$.  The curves in the latter plots
were not overly sensitive to $\sigma$ so we used the recent value
 $\sigma=0.4742$ \cite{Gracey2015}.}

For the BCC and FCC lattices, we find significantly more precise values of $p_{c}$ than van der Marck \cite{vanderMarck98}, who gave only two digits of accuracy. And we give for the first time a value of $p_{c}$ for the SC-NN+2NN lattice, which was not studied before for bond percolation. 

\begin{table}[htb]
\caption{Estimations of bond percolation thresholds for the 4D percolation models.}
\begin{tabular}{c|c|c|c}
\hline\hline
    lattice  & $z$ & $p_{c}$ (present) & $p_{c}$ (previous) \\ \hline
    SC       & 8   & 0.1601312(2)      & 0.16005(15) \cite{AdlerMeirAharonyHarrisKlein90}\\
             &     &                   & 0.160130(3) \cite{PaulZiffStanley2001}\\ 
             &     &                   & 0.1601314(13) \cite{Grassberger03}\\
             &     &                   & 0.1601310(10) \cite{DammerHinrichsen2004}\\
             &     &                   & 0.16013122(6) \cite{MertensMoore2018}\\
    BCC      & 16  & 0.074212(1)       & 0.074(1) \cite{vanderMarck98} \\
    FCC      & 24  & 0.049517(1)       & 0.049(1) \cite{vanderMarck98} \\
    SC-NN+2NN & 32  & 0.035827(1)       & ----- \\
\hline\hline
\end{tabular}
\label{tab:ept}
\end{table}

\begin{figure}[htbp] 
\centering
\includegraphics[width=3.8in]{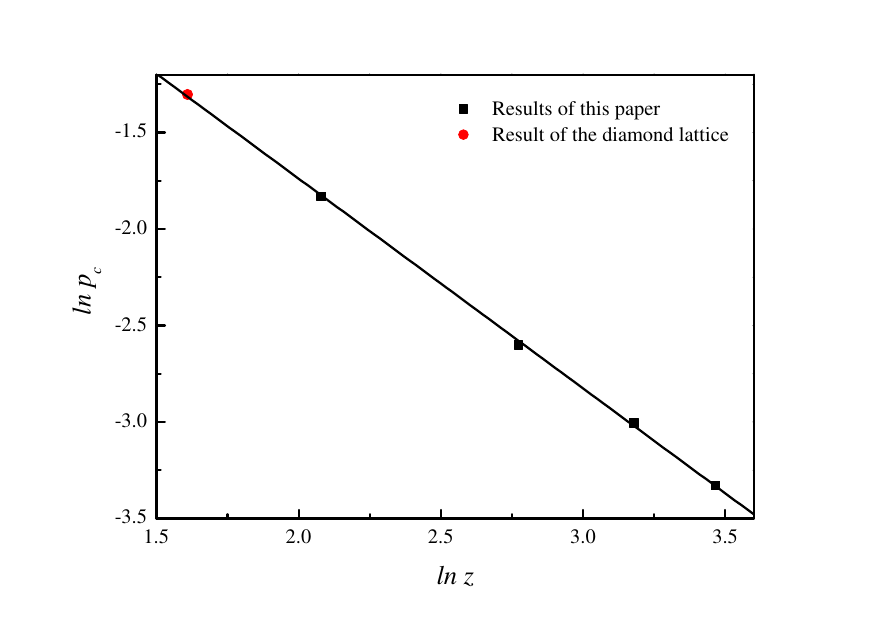} 
\caption{A log-log plot of percolation thresholds $p_{c}$ vs.\ coordination number $z$ for the lattices simulated in this paper (square symbols) and the diamond lattice (circle) provided in Ref.\  \cite{vanderMarck98}. The slope gives an exponent of $a = 1.087$ in Eq.\ (\ref{eq:scaling}), and the intercept of the line is at $\ln p_c = 0.435$. Also shown on the plot are the site thresholds for the same five
lattices (triangles) \cite{vanderMarck98,KotwicaGronekMalarz19,MertensMoore2018}, in which case the linearity of the data is not nearly as good.}
\label{fig:ln-pc-z-gamma}
\end{figure}

Table \ref{tab:ept} also shows the coordination number $z$ for each lattice. The values of $p_{c}$ decrease with the coordination number $z$ as one would expect.  
Finding correlations between percolation thresholds and lattice properties has a long history in percolation studies \cite{ScherZallen70,vanderMarck97,Wierman02,WiermanNaor05}. In Ref.\ \cite{KurzawskiMalarz2012} it was found that the site thresholds for several 3D lattices can be fitted by a simple power-law in the coordination number $z$
\begin{equation}
p_{c}(z) \sim z^{-a},
\label{eq:scaling}
\end{equation}
with $a =  0.790(26)$ in 3D.  Similar power-law relations for various systems were studied by Galam and Mauger \cite{GalamMauger96}, van der Marck \cite{vanderMarck98}, and others, usually in terms of  $(z-1)^{-a}$ rather than vs.\ $z^{-a}$.  Making a log-log plot of the 4D data of Table \ref{tab:ept}, along with the bond threshold $p_c = 0.2715(3)$ for the 4D diamond lattice \cite{vanderMarck98}, which has coordination number $z=5$,  in Fig.\ \ref{fig:ln-pc-z-gamma}, we find $a = 1.087$.  Deviations of the thresholds from this line are within about 2\%.   We note that the data for site percolation thresholds of these lattices, taken from \cite{vanderMarck98,KotwicaGronekMalarz19,MertensMoore2018}, do not show such a nice linear behavior as do the bond thresholds, as shown in Fig.\ \ref{fig:ln-pc-z-gamma}.   \red{We do not know any reason for this excellent power-law dependence of the bond thresholds, nor why the exponent has the value of approximately 1.087.}

\section{Conclusions}
In this paper, by employing the  single-cluster growth algorithm, bond percolation on SC, SC-NN+2NN, BCC, and FCC lattices in 4D was investigated. The  algorithm allowed us to estimate the percolation thresholds with high precision with a moderate amount of calculation. For the BCC and FCC lattices, our results are about three orders of magnitude more precise than previous values, and for SC-NN+2NN lattice, we find a value of the bond percolation threshold for the first time. In addition, the results indicate that the percolation thresholds $p_{c}$ decrease monotonically with the coordination number $z$, quite accurately according to a power law of $p_{c} \sim z^{-a}$, with the exponent $a = 1.087$.

There remain many lattices where thresholds are not known, or where they are known only to \red{low significance, such as two or three digits}, and the methods described here can be used to find them with high accuracy in a straightforward manner.  For example, the bond thresholds on the many complex neighborhood lattices of Malarz and co-workers have not been determined before, and knowing these thresholds may be useful for various applications.

Another result of this paper was a precise measurement of the exponent $\tau$, which we were able to do using the finite-size scaling behavior of Eq.\ (\ref{localslope}), which requires the knowledge of $\Omega$ although the results for $\tau$ are not very sensitive to the precise value of $\Omega$.  Averaging the results over the four lattices, we find $\tau = 2.3135(5)$.  This is consistent with previous Monte Carlo values of $2.3127(6)$  \cite{BallesterosEtAl97}, 2.313(3) \cite{PaulZiffStanley2001}, 2.313(2) \cite{Tiggemann01}, the recent Monte Carlo result of Mertens and Moore, 2.3142(5) \cite{MertensMoore2018}, and also close to the recent four-loop series result 2.3124 of Gracey \cite{Gracey2015}.  In concurrent work, Deng et al.\ find that the fractal dimension in 4D equals $d_f = 3.0446(7)$, which implies by the scaling relation $\tau = 1 + d/d_f = 2.3138(3)$ \cite{DengEtAl2019}. Our value 2.3135(5) is a good average of all these measurements.  

We have also found a fairly accurate value of the corrections-to-scaling exponent $\Omega$, with the result $0.40(3)$, which also gives a value of $\omega = \Omega d_f = 1.22(9)$.  We determined $\Omega$ by adjusting its value until we found a straight line in plots like Figs.\ \ref{fig:sc-localslope-vs-s-omega} and \ref{fig:sc-s-tau-2-Ps-vs-s-omega} --- while simultaneously trying to find $p_c$ and $\tau$.  Having three different kinds of plots for each lattice helped in this simultaneous determination of these three parameters.  Previous Monte-Carlo values of $\Omega$ were 0.31(5) \cite{AdlerMeirAharonyHarris90}, 0.37(4) \cite{BallesterosEtAl97}, and 0.5(1) \cite{Tiggemann01}.  In Ref.\ \cite{Gracey2015}, Gracey gives the series extrapolation of $\Omega = 0.4008$ \cite{Gracey2015}, which was based upon a Pad\'e approximation assuming the value of $\Omega = 2$ for 2D.  Redoing that calculation using $\Omega = 72/91$ (2D) from Refs.\ \cite{AharonyAsikainen03,Ziff11b}, Gracey finds $\Omega = 0.3773$  \cite{Gracey19}.  Both of these values (0.4008 and 0.3773) are  consistent with our result of $\Omega = 0.40(3)$.  

In forthcoming papers \cite{XunZiff20,XunZiff20b} the authors will report on a study of many 3D lattices with complex neighborhoods for both site and bond percolation.

\section{Acknowledgments}
We are grateful to the Advanced Analysis and Computation Center of the China University of Mining and Technology for the award of CPU hours to accomplish this work. This work is supported by the China Scholarship Council Project (Grant No. 201806425025) and the National Natural Science Foundation of China (Grant No.\ 51704293).

\bibliographystyle{unsrt}
\bibliography{bibliography.bib}

\end{document}